\documentclass[preprint,authoryear,3p]{elsarticle}

\usepackage{lineno,hyperref}
\modulolinenumbers[5]
\usepackage{amsmath}
\usepackage{graphicx}
\usepackage{dcolumn}
\usepackage{bm}
\usepackage{color}

\definecolor{orange}{rgb}{1.0, 0.65, 0.0}
\definecolor{brown}{rgb}{0.6, 0.46, 0.33}

\journal{Journal of Informetrics}

\biboptions{authoryear}

\begin{document}

\begin{frontmatter}

\title{Investigating the interplay between fundamentals of national research systems: performance, investments and international collaborations}

\author[mymainaddress,mysecondaryaddress]{Giulio Cimini\corref{mycorrespondingauthor}}
\author[mysecondaryaddress]{Andrea Zaccaria}
\author[mysecondaryaddress,mymainaddress]{Andrea Gabrielli}
\cortext[mycorrespondingauthor]{Corresponding author (giulio.cimini@roma1.infn.it)}
\address[mymainaddress]{IMT - Institute for Advanced Studies, Piazza San Ponziano 6, 55100 - Lucca, Italy}
\address[mysecondaryaddress]{Istituto dei Sistemi Complessi (ISC)-CNR, UoS Dipartimento di Fisica, Universit\`a ``Sapienza'', Piazzale Aldo Moro 5, 00185 - Rome, Italy}

\begin{abstract}
We discuss, at the macro-level of nations, the contribution of research funding and rate of international collaboration to research performance, 
with important implications for the ``science of science policy''. In particular, we cross-correlate suitable measures of these quantities with a scientometric-based assessment of scientific success, 
studying both the average performance of nations and their temporal dynamics in the space defined by these variables during the last decade. 
We find significant differences among nations in terms of efficiency in turning (financial) input into bibliometrically measurable output, 
and we confirm that growth of international collaboration positively correlate with scientific success---with significant benefits brought by EU integration policies. 
Various geo-cultural clusters of nations naturally emerge from our analysis. We critically discuss the factors that potentially determine the observed patterns.
\end{abstract}

\begin{keyword}
Science of Science policy\sep scientific impact \sep R\&D funding \sep international collaborations
\end{keyword}

\end{frontmatter}

%\linenumbers

%%%%%%%%%%%%%%%%%%%%%%%%%%%%%%%%%%%%%%%%%%%%%%%%%%%%%%%%%%%%%%%%%%%%%%%%%%%%%%%%%%%%

\section{Introduction}\label{intro}

The science of science policy is emerging as an interdisciplinary field that aims at developing theoretical models and studying empirical evidence 
for the performance of scientific communities and individual researchers \citep{ScioSci}. This scientific activity can then help to develop policies 
for improving Research and Development (R\&D) funding allocation and strategical decision making. Within the field, a critical issue 
has been that of identifying suitable quantities to characterize the research systems at the level of nations, in terms of scientific impact, development and competitiveness.

Indeed, many metrics to evaluate the impact of scientific research have been proposed in the literature, but few have proven to be satisfactory---see \citet{Waltman_rev} 
for a recent overview of the field. The traditional approach, based on shares of citations or documents \citep{May1997,King2004}, in fact, suffers from several drawbacks. 
First, the number of published papers gives no clear information about the quality of the research they contain. Second, the number of published documents grows steadily in time, whereas, 
citation statistics are highly biased for recent papers that had not enough time to attract citations \citep{Medo2011}, and thus need to be normalized properly for a time dynamical analysis. 
Third, the number of citations or documents is an extensive measure that naturally correlates with size, thus requiring additional normalization in order to compare, e.g., different national research systems. 
The latter problem applies also to more refined methods like the H-index \citep{Hirsch2005} and its variants. 
Other approaches \citep{Smith2014} measure scientific performance of individual papers by comparing the total number of citations a paper has accrued 
to those of other publications of the same journal volume. Still, methods based on publication venues suffer from all the exogenous and endogenous factors 
that enter in the effective publication mechanism and that can follow different criteria than the real quality of the scientific work. 

Metrics that take care of the skewness of citation distributions (by considering only highly cited publications) \citep{Aksnes2004} have found wide application in the field, 
however how to determine whether a publication is counted as highly cited or not is still an open issue \citep{Waltman2013} 
which can hinder a comparison of different studies \citep{Bornmann2013}. 
In order to avoid all the problems mentioned above, and to obtain a proper normalization of bibliometric data, we follow the general ideas of \citet{Waltman2011} 
and measure scientific performance of individual nations as their ratio of citation shares to publication shares (see Section 2 below). 
The reason is that whenever a nation receives a larger share of citations compared to the fraction of papers it publishes, it is producing science that has a greater impact than the world average. 

Interestingly, most national research systems have been characterized, during the last years, by a remarkable increase of the number of international scientific collaborations \citep{Leydesdorff2008,Leydesdorff2013}. 
This phenomenon has been studied and analyzed especially in the context of the European Union, where it appears to be a particularly strong clue of successful EU integration policies---see 
\citet{Glanzel2007,Huang2011}; and \citet{Chessa2013} for a contrary view. However, also developing nations have increased their rates of collaboration with foreign, already developed nations, 
and empirical evidence suggests that this strategy is at the core of their successful entrance in the scientific community \citep{Wagner2001}. As noted by \citet{Persson2010}, 
it is necessary to point out that the presence of a possible cause-effect relationship between scientific success and international collaborations is still an open issue. 
Notably, simple but commonly adopted measures of scientific performance (productivity, citation performance and journal placement) 
are known to be positively correlated with the rate of internationalization of the scientific community of a nation \citep{Katz1997,Abramo2011,Kato2013,Smith2014}. 
In particular, it has been shown that the most successful teams are characterized by a moderate level of cultural diversity \citep{Barjak2008}.

Of course, any study of national scientific performance cannot neglect the role played by the availability of financial resources---namely, R\&D funding. 
Yet, assessing efficiency at the research system level is a complex research question. In a recent paper, \citet{Pan2012} 
have shown that the research impact of a nation grows linearly with the amount of national R\&D funding, pointing out also the presence of a peculiar effect: 
in order to be effective, public investments should exceed a certain threshold. As pointed out by \citet{Leydesdorff2009}, there is a great difference in national ability 
to transform financial input into bibliometric output. The situation becomes even more complicated when looking at scales smaller than nations. 
For instance, according to the analyses performed by \citet{Sandstrom2014} comparing the change in scientific output with the change of funding, 
there is no evidence that the amount of institutional funding correlates with competitiveness, overall performance, and top performance of universities at the national level. 
\citet{Fortin2013} instead focused on individual researchers, showing that impact is positively, but only weakly, related to funding, 
and in general is a decelerating function of funding itself. These conclusions, together with the multi-facet structure of the R\&D funding scheme \citep{Leydesdorff2009}, 
stress the need of a systematic approach to funding-based analysis.

Notably, as we show at the end of this paper, a complex structure of geo-cultural clusters naturally emerges from this kind of studies. 
As originally pointed out by \citet{Frame1979}, international co-authorships are clearly biased by extra-scientific factors such as geography, politics and language. 
Also \citet{Luukkonen1992} reached similar conclusions, suggesting the presence of cultural \textit{centers} on which other nations hinge. 
In summary, three fundamental aspects naturally emerge as prominent features for a systematic analysis of nations scientific production: 
internationalization, funding, success, and, as a further resulting output, the presence of geographic and cultural communities. 
In this work we precisely address the issue of how the complex interaction between these fundamentals shape the scientific production of nations. %and their possible success. 

Our paper is organized as follows. In Section 2 we describe our datasets and define the variables we are going to use in our analysis. 
In Section 3 we present our main results, namely, a static and dynamic analysis for the scientific performance of nations as a function of both level of internationalization 
and fundings to various types of research institutions. The concluding Section 4 summarizes our findings and discusses future perspectives.

\section{Materials and Methods}\label{sec:method}

In this section we define the different metrics we rely upon to characterize national research systems, and describe the databases used to build them. 

\subsection{OECD data and R\&D funding}

We collect data on national expenditure in scientific research and development from the Organization for Economic Cooperation and Development (OECD, www.oecd.org). 
Data refer to $N_f=37$ developed nations and to years 2000-2012. All expenditures are expressed in terms of current purchasing power parity (in millions of US dollars). 
The overall national expenditure indicator, known as GERD (Gross Expenditure on R\&D), is divided into three main components: 
BERD (Business Expenditure on R\&D), namely R\&D expenditure performed in the business sector,~\footnote{The business sector includes firms, organizations and institutions 
whose primary activity is the market production of goods or services (other than higher education), and the private non-profit institutions mainly serving them.} including both public and private fundings; 
HERD (Higher Education Expenditure on R\&D), expenditures for basic research performed in the higher education sector,~\footnote{The higher education sector includes universities, 
colleges of technology and other institutions of post-secondary education, and the research institutes, experimental stations and clinics operating 
under the direct control of, administered by or associated with higher education institutions.} again including both public and private fundings; 
and GOVERD (Government Intramural Expenditure on R\&D), expenditures in the government sector~\footnote{The government sector includes departments, offices and other bodies 
which furnish (but normally do not sell) to the community common services other than higher education, as well as those that administer the state and the economic and social policy of the community, 
and the non-profit institutions controlled and mainly financed by government but not administered by the higher education sector.} (we refer to \citet{OECD2002} for a more detailed definition of these quantities).
BERD is arguably important for innovation and economic growth, being closely linked to the creation of new products and production techniques (patents), 
and at the end to the innovation efforts of a nation. Thus, in the context of studies focused on bibliometric scientific outputs (namely, papers), usually only HERD and GOVERD are 
considered to be relevant. In particular, Leydesdorff and Wagner \citep{Leydesdorff2009} pointed out that HERD cannot be considered as a sufficient indicator of input to academic research, 
because in some nations (like China and Russia) GOVERD becomes larger than HERD. However, they also noted that the public research sector is often mission-oriented and therefore less driven 
by the institutional and scientific need to publish \citep{OECD2002}. Since a shared consensus on what kind of input should be considered (to relate scientific success with) is still missing, 
here we take into account all three indicators (BERD, HERD and GOVERD) separately. Note that since we are interested in assessing the quality of the scientific output of a nation, 
we consider intensive metrics, that is, size-independent quantities obtained by normalization with the respective nation GDP. 
We denote as $\mathcal{B}_i(t)$, $\mathcal{H}_i(t)$ and $\mathcal{G}_i(t)$, respectively, such normalized BERD, HERD and GOVERD values for nation $i$ during year 
$t$.~\footnote{In order to compensate for the few missing (yearly) values in the database, we used linear interpolation based on the available data.}

\subsection{SCImago data, Impact and Internationalization}

In order to measure the impact of scientific output and its level of internationalization, 
we use bibliometric data collected from the SCImago website (www.scimagojr.com)---aggregated from the Scopus database (www.scopus.com). 
Data refer to $N_d=239$ nations, $D=28$ scientific domains and $d=311$ scientific sub-domains (each belonging to one domain), and cover years 1996-2013. 
SCImago provides basic statistics on national scientific output: $d_{i\alpha}(t)$, the number of scientific documents a nation $i$ published on domain $\alpha$ 
during year $t$; $c_{i\alpha}(t)$, the number of citations accrued by those papers from $t$ up to now; and $d_{i\alpha}^*(t)$, the number of documents 
published by nation $i$ on domain $\alpha$ during year $t$ whose affiliations include at least another nation address. 
These quantities are obtained aggregating data of individual papers from the whole corpus of scientific literature, and using a full counting approach. 
We remand the reader to section 2.2.3 below for more details on these points.

\subsubsection{Bibliometric impact}

We use the basic SCImago statistics to first compute the average (or expected) citations of documents published in year $t$ and in domain $\alpha$, 
defined as $e_\alpha(t)=[\sum_jc_{j\alpha}(t)]/[\sum_jd_{j\alpha}(t)]$. 
Then, we assess the scientific impact (or success) of nation $i$ during year $t$ through either the $CPP/FCSm$ indicator (citations per publication over mean field citation score) 
or the $MNCS$ indicator (mean normalized citation score) \citep{Waltman2011}, which in our notation are formalized as:
\begin{equation}
 [CPP/FCSm]_i(t)=\frac{\sum_\beta c_{i\beta}(t)}{\sum_\beta e_\beta(t)\,d_{i\beta}(t)}=\left(\sum_\beta c_{i\beta}(t)\right) \Bigg/ \left(\sum_\beta d_{i\beta}(t)\frac{\sum_jc_{j\beta}(t)}{\sum_jd_{j\beta}(t)}\right),
 \label{eq.CPP_FCSm}
\end{equation}
\begin{equation}
 [MNCS]_i(t)=\frac{\sum_\beta [c_{i\beta}(t)/e_\beta(t)]}{\sum_\beta d_{i\beta}(t)}=\left(\sum_\beta c_{i\beta}(t)\frac{\sum_j d_{j\beta}(t)}{\sum_j c_{j\beta}(t)} \right) \Bigg/ \sum_\beta d_{i\beta}(t).
 \label{eq.MNCS}
\end{equation}
$CPP/FCSm$ is built by dividing the total citations obtained by a nation with the expected number of citations it should have received for its publications, whereas, 
$MNCS$ is the average of the ratio of actual citations to expected citations for each publication. Note that while both metrics are intensive and thus apt for comparing research systems of different scales, 
they are inherently different. In particular, the $MNCS$ indicator includes field-specific normalization, which may be necessary as nations can concentrate their scientific efforts into different fields. 
However, an agreement on which of these variants should be preferred is still lacking in the literature \citep{Waltman_rev}. 
Here, we add to the discussion by exploring two variants of these indices that, in our opinion, are equally intuitive and can provide different insights. 
Our first proposal consists in normalizing, for a given nation, its share of world citations with its share of world documents. This translates into:
\begin{equation}
 [Csh/Dsh]_i(t)=\left(\frac{\sum_\beta c_{i\beta}(t)}{\sum_\beta \sum_j c_{j\beta}(t)}\right) \Biggm/ \left(\frac{\sum_\beta d_{i\beta}(t)}{\sum_\beta \sum_j d_{j\beta}(t)}\right).
 \label{eq.success}
\end{equation}
Of course, such ratio of shares can be also measured within a given domain $\alpha$, and then averaged over all domains---an approach that brings to the following alternative measure of overall success:
\begin{equation}
[MNCsh]_i(t)=\frac{1}{D}\sum_\beta\left(\frac{c_{i\beta}(t)}{\sum_j c_{j\beta}(t)}\right) \Bigg/ \left(\frac{d_{i\beta}(t)}{\sum_j d_{j\beta}(t)}\right).
 \label{eq.success_single}
\end{equation}
Again, the difference between $Csh/Dsh$ and $MNCsh$ resides in how the different scientific domains are weighted in the averaging procedure---with $Csh/Dsh$ resembling $CPP/FCSm$ 
and $MNCsh$ closer in spirit to $MNCS$. In particular, $Csh/Dsh$ weights all papers equally, thus it does not distinguish between documents belonging to different scientific areas, 
whereas, $MNCsh$ weights the all scientific fields equally---meaning that it is field-normalized like the $MNCS$ indicator. 
However, empirical comparisons between all these approaches \citep{Waltman2011} show that the differences are small, 
especially at the level of nations (see also Table \ref{tab.corr} for a correlation analysis between the different indicators). 
Thus, to our purpose these metrics are interchangeable. In this work, we have chosen to use $Csh/Dsh$ as this index 
is, among those considered here, the least subject to noisy fluctuations that affect domains with overall few documents and citations. 
Additionally, it is the only one that is independent on the specific classification used for scientific sectors, an aspect that is particularly relevant. In fact, 
many researchers have raised important issues related to the choice of a classification system \citep{Waltman_rev}, also in relation 
to the ever increasing amount of interdisciplinary research papers. Yet, by using $Csh/Dsh$ we pay the price of loosing a proper field-normalization. 
In the following, we will denote the scientific success of nation $i$ during year $t$ as $\mathcal{S}_i(t)$.

\begin{table}[h!]
\begin{center}
\begin{tabular}{l|c}
$Csh/Dsh - CPP/FCSm$ & 0.98 \\
$Csh/Dsh - MNCS$ & 0.96 \\
$Csh/Dsh - MNCsh$ & 0.97 \\
$MNCsh - CPP/FCSm$ & 0.99 \\
$MNCsh - MNCS$ & 0.99 \\
$CPP/FCSm - MNCS$ & 0.99 \\
\end{tabular}
\end{center}
\caption{Pearson correlation coefficients between the four indices of success. 
As detailedly explained in Section 3, impact metrics are computed for 46 developed countries as the time average of year-specific indicators from 2004 to 2012.}
\label{tab.corr}
\end{table}

\subsubsection{Internationalization}

To quantify the level of internationalization of the research system of a nation, we use the rate of international collaborations that, for a given nation $i$ in domain $\alpha$ during year $t$, 
is defined as $\mathcal{I}_i(t)=[\sum_\beta d_{i\beta}^*(t)]/[\sum_\beta d_{i\beta}(t)]$. 
Internationalization can be also measured within a given domain $\alpha$ as $\iota_{i\alpha}(t)=d_{i\alpha}^*(t)/d_{i\alpha}(t)$, 
which then brings to an alternative metric for overall internationalization: $\mathcal{I}'_i(t)=\sum_{\alpha}\iota_{i\alpha}(t)/D$. The two approaches lead to very similar quantitative results. 
According to the same reasoning used for the choice of the impact indicator, in the following we use the first definition of internationalization of national publication baskets. 

\subsubsection{Remarks}

Before proceeding to results, let us point out some technical issues concerning the use of the SCImago dataset and the consequent methodological restrictions imposed.

{\em Social Sciences and Humanities} (SSH) {\em and Scopus coverage} --- Scopus (and other bibliometric databases as well) have an almost complete coverage of documents 
written in English and published in international peer-reviewed journals, whereas, documents written in languages other than English and published in national journals are not covered in full---also 
if they have a significant share in the database. In this situation, the most penalized branches appear to be the SSH \citep{Sivertsen2012}. In fact, research in SSH has a number of peculiar features 
with respect to other fields: it has a stronger national and regional orientation, is published less in journals and more in books, has a slower pace of theoretical development, is less collaborative 
and is directed more at a non-scholarly public \citep{Nederhof2006}. As SCImago data are aggregated spearately for each scientific sector, the SSH domains could have been excluded 
from our analysis. For the sake of completeness, we decided otherwise; yet, it is important to keep in mind that, while the SSH are few and their weight is thus small, 
including them could result in a slight bias towards anglophone nations---that, as we will see, may be the reason for the slightly better performance of Commonwealth members with respect to Western Europe. 
Another related issue is whether to consider all publications appearing in the Scopus database, or only publications in international scientific journals, 
i.e., basically {\em core} publications (see \url{http://www.leidenranking.com/methodology/indicators\#core-publications} for a definition of core publication). 
In this respect, the SCImago statistics are built following the former approach which, remarkably, leads to the highest coverage by including also 
non-core papers---that may be as scientifically relevant as core ones. The alternative approach, however, seems useful especially at small scales 
(e.g., at the level of individual universities) by leading to more accurate impact indicators \citep{Waltman_rev}.

{\em Skewness of the citation distribution} --- Citation distributions are extremely skewed. As a consequence, average-based indicators (like those we use here) 
can be sensitive to the presence of one or a few very highly cited publications \citep{Aksnes2004}. 
Percentile-based indicators \citep{Waltman2013} are less sensitive to these outliers, and are thus natural candidates for measuring scientific success. 
Despite the many advantages of these indicators \citep{Bornmann2013}, in this work we do not rely on them as they cannot be extracted from the available SCImago statistics. 
Yet, this is not a crucial issue when considering large nations, large aggregation levels to determine scientific areas, and wide temporal windows---as we do in this study, and differently from 
\citet{Aksnes2004}. The reason is the law of large numbers \citep{feller2008introduction}, which allows to assume safely in this case that distortions potentially affecting 
a single paper are smoothed out. This means that citation distributions are indeed skewed but their tails are well-defined (as we have enough statistics for them), 
and if they are not too broad (so that the mean is not divergent), average-based indicators can be reliably used as well to measure success.

{\em Counting method} --- In principle, impact indicators can be calculated using either a full counting or a fractional counting method, 
two alternative ways to assign internationally co-authored publications to countries \citep{Waltman_rev}. 
In the full counting approach, a publication co-authored by various countries is fully assigned to each of the these countries. 
Fractional counting instead assigns a publication to a country with a weight, proportional for instance to the fraction of authors or affiliations in the publication belonging to that country. 
The full counting method can thus be seen as measuring participation, while the fractional counting as measuring how many papers are creditable to a country \citep{Aksnes2012}.
In this respect, collaboration indicators (e.g., the amount of international collaborations) are always calculated using the full counting method. 
Concerning citation indicators, full counting is also commonly adopted. However, recent works \citep{Aksnes2012,Waltman2015} have argued that fractional counting 
is the correct approach to use for country-level analyses, as it leads to a proper field normalization of impact indicators---and therefore to a more fair comparisons between fields, 
and between countries active in different fields. Yet, the use of fractional counting is complicated by the several ways in which weights can be assigned \citep{Waltman2015}, 
and by the additional possibility to use half-way counting methods \citep{Smith2014}. 
In this work, full counting is adopted as the SCImago statistics are built according to this criterion. 
Nevertheless, we can point out qualitatively the expected differences in outcome, were we given the possibility to chose fractional counting. 
The main observation is that larger countries have, in general, a lower degree of international co-authorship among their publication output, 
and are thus penalized by full counting with respect to small countries with high level of internationalization---as shown by \citet{Aksnes2012}. 
Thus, as we will see, the fact that in our study European nations perform relatively better than, e.g., the United States, Russia or China, 
is partly due to the counting method adopted. Yet, since the difference between the two methods basically consists in a country-specific rescaling of impact indicators, 
the relative temporal changes of countries impact we will analyse are, {\em per se}, unaffected by the choice of the counting method.

\section{Results and Discussion}\label{sec:results}

We now present and discuss the results of our analysis on the complex relationships between the three fundamental features of national research systems defined above 
(funding, success and internationalization). We focus over the years 2004-2012, for which we have the maximum data coverage,\footnote{We excluded years up to 2003 as we observe 
a systematic drop of international collaborations during years 2001-2003 for all nations, whose origin remains unclear to us also after a discussion with the administrators of SCImago.} 
and present two kinds of analysis: (I) the study of interdependences between the mean values of these quantities over the considered time window of $T=9$ years, and 
(II) the study of the temporal evolution (i.e., of the trajectories) of nations in the space of the fundamentals over this time span. 
Time averages are denoted with a bar over the respective symbol; we thus have, for a given nation $i$, $\bar{\mathcal{S}}_i=\frac{1}{T}\sum_t \mathcal{S}_i(t)$, 
$\bar{\mathcal{I}}_i=\frac{1}{T}\sum_t \mathcal{I}_i(t)$, $\bar{\mathcal{B}}_i=\frac{1}{T}\sum_t \mathcal{B}_i(t)$, $\bar{\mathcal{H}}_i=\frac{1}{T}\sum_t \mathcal{H}_i(t)$ 
and $\bar{\mathcal{G}}_i=\frac{1}{T}\sum_t \mathcal{G}_i(t)$ respectively as its average success (or impact), internationalization, BERD/GDP, HERD/GDP and GOVERD/GDP over the considered time window. 
For the sake of graph readability, we use different colors to plot averages and trajectories of $N=46$ nations according to the following classification 
based on geographical, historical and cultural factors:\footnote{R\&D expenditure data for Brazil, Chile, Egypt, Hong Kong, India, Iran, Malaysia, Thailand and Ukraine 
(as well as BERD data for New Zealand, South Africa and Switzerland) are missing from the OECD database, 
and thus these countries are missing from plots related to research funding.}\\
\begin{itemize}
 \item \textcolor{black}{\bf Black}: United States;
 \vspace{-0.2cm}
 \item \textcolor{blue}{\bf Blue}: Commonwealth (Australia, Canada, Hong Kong, India, Malaysia, New Zealand, Singapore, South Africa, United Kingdom);
 \vspace{-0.2cm}
 \item \textcolor{green}{\bf Green}: Western and Southern Europe (Austria, France, Germany, Greece, Ireland, Italy, Portugal, Spain);
 \vspace{-0.2cm}
 \item \textcolor{cyan}{\bf Cyan}: Switzerland and Northern Europe (Belgium, Denmark, Finland, Netherlands, Norway, Sweden);
 \vspace{-0.2cm}
 \item \textcolor{red}{\bf Red}: Eastern Europe (Czech Republic, Hungary, Poland, Romania, Russian Federation, Slovakia, Slovenia, Ukraine);
 \vspace{-0.2cm}
 \item \textcolor{brown}{\bf Brown}: Middle East (Egypt, Israel, Iran, Turkey);
 \vspace{-0.2cm}
 \item \textcolor{orange}{\bf Orange}: Asian region (China, Japan, South Korea, Thailand, Taiwan);
 \vspace{-0.2cm}
 \item \textcolor{magenta}{\bf Magenta}: Latin America (Argentina, Brazil, Chile, Mexico).
\end{itemize}

\subsection{The internationalization of scientific research}

\begin{figure}[h!]
\begin{center}
\includegraphics[width=0.8\textwidth]{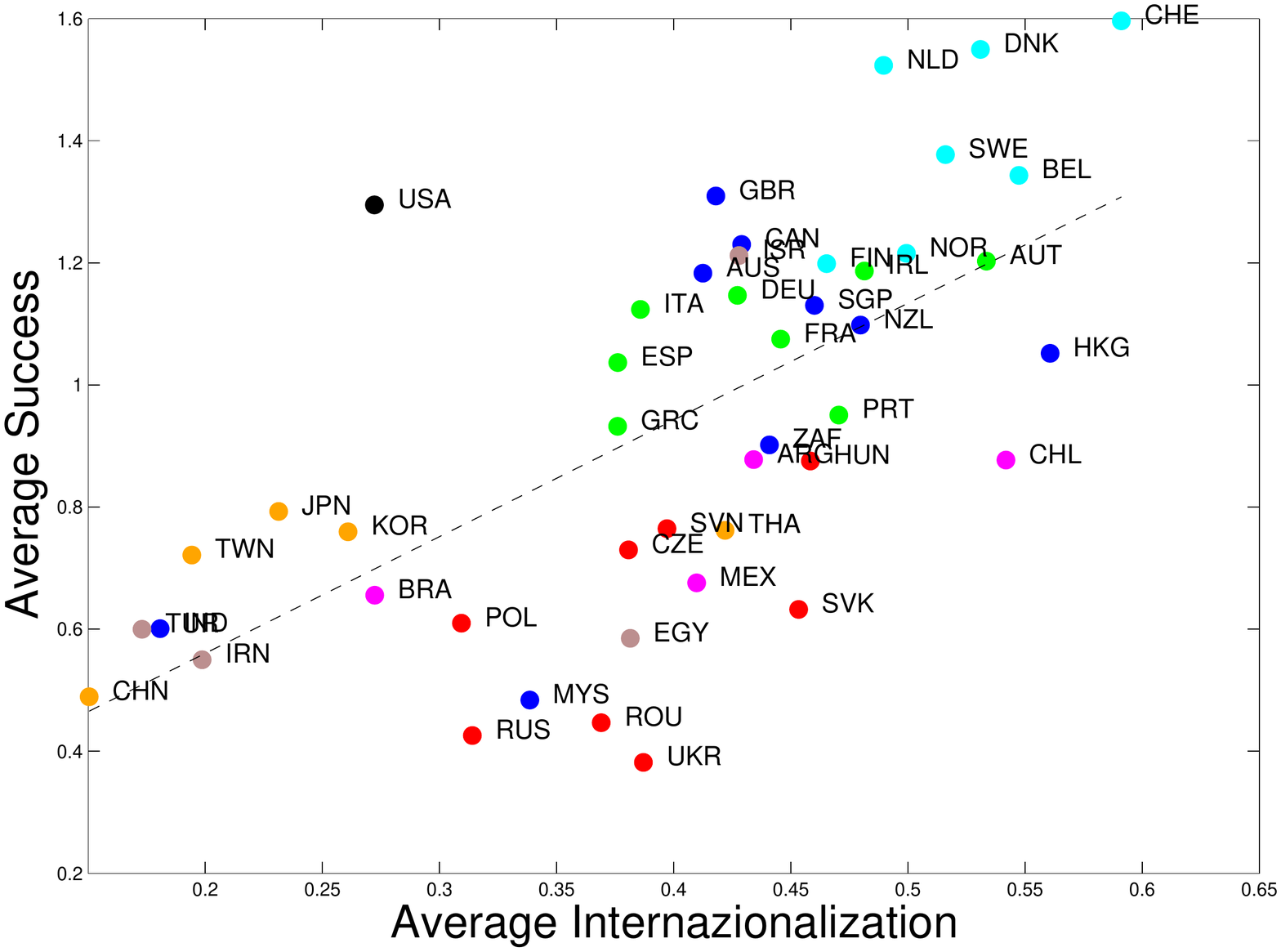}
\end{center}
\vspace{-0.5cm}
\caption{Relation between average internationalization $\bar{\mathcal{I}}$ and average scientific success $\bar{\mathcal{S}}$ of nations. 
Linear regression of data with slope $1.91\pm0.33$, intercept $0.18\pm0.14$ and $R^2=0.43$ is shown as a dashed segment.}
\label{fig:SuccINT}
\end{figure}

\begin{figure}[h!]
\begin{center}
\includegraphics[width=0.8\textwidth]{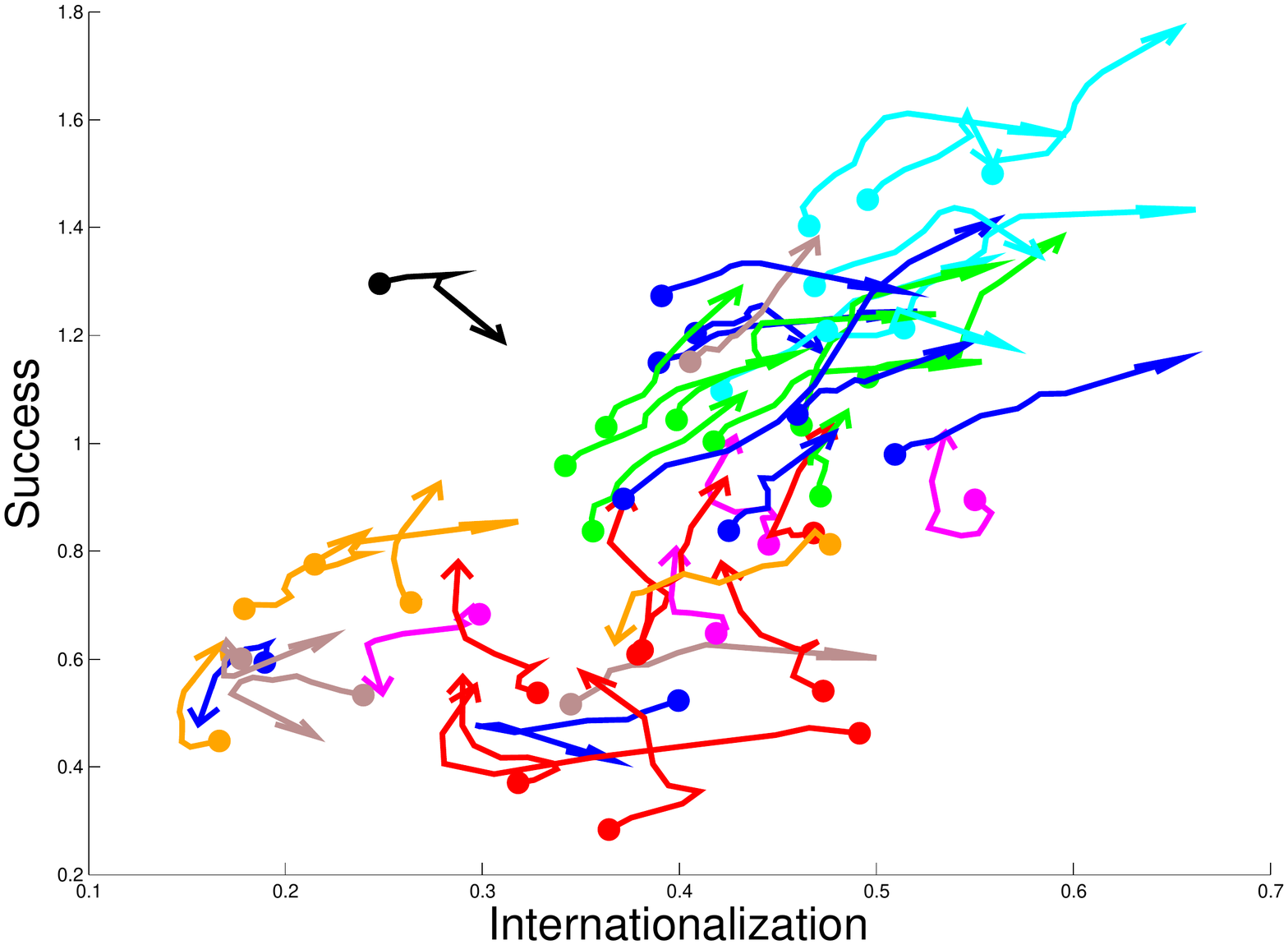}
\end{center}
\vspace{-0.5cm}
\caption{Temporal evolution of nations in the plane of international collaboration $\mathcal{I}(t)$ and scientific success $\mathcal{S}(t)$.}
\label{fig:SuccINT_evol}
\end{figure}

Figures \ref{fig:SuccINT} and \ref{fig:SuccINT_evol} report results of our analysis for the relation between internationalization and scientific success. 
Looking at the average values of these quantities (figure \ref{fig:SuccINT}), we notice that they are significantly correlated, with a main trend that starts 
from Eastern European nations (lowest $\bar{\mathcal{I}}$ and $\bar{\mathcal{S}}$) and ends with Northern European nations and Switzerland (highest $\bar{\mathcal{I}}$ and $\bar{\mathcal{S}}$), 
with Western Europe and Commonwealth members in between. Significant outliers are represented by the United States that, because of their large self­-consistency with respect to any other nation, 
rely much less on international collaboration to achieve the same level of scientific performance of the other western countries, and by Asian countries whose research systems are the least internationalized 
(supposedly for linguistic and cultural reasons). An even more interesting picture emerges from the time dynamics analysis of nations in the $\mathcal{I}-\mathcal{S}$ plane (figure \ref{fig:SuccINT_evol}). 
For low values of $\mathcal{I}$ and $\mathcal{S}$ (mostly Asian and Middle East countries) we observe a chaotic-like dynamics, which however becomes more laminar as soon as the amount of international collaboration 
increases slightly. There we observe the Eastern European countries trying to catch up with the group of developed nations in scientific success, without relying much on increasing the rate of internationalization. 
Then, once scientific success approaches values close to 1, basically every nation enters into a stream that allows to increase success even more by increasing the amount of international collaborations. 
Notably, such a region comprises mostly Western and Northern European countries, where internationalization is stronger supposedly because of EU research integration policies. 
These policies thus seem to be rather beneficial in improving national research performances. Finally, United States are located away from the main stream (but are directed towards it), 
meaning that the increasing internationalization of their research system is not leading to immediate performance improvements---which will perhaps occur once they reach values of $\mathcal{I}$ 
similar to those of the other western nations. Note also that the slightly decreasing success of United States is also caused by the increasing success of almost all other nations: 
since our measure of success is based on citation shares and document shares, and the world's share is necessarily equal to one, 
if the share of a country increases then the share of one or more other countries is forced to decrease.

We conclude this section with the following key message: Internationalization emerges as a fundamental parameter for the scientific development of nations. 
In this respect, the European Union mission of promoting integration among its constituent nations appears to be well-founded---yet, more for old members than for Eastern European countries.

\subsection{Outcome of R\&D investments}

\begin{figure}[p]
\begin{center}
\includegraphics[width=0.8\textwidth]{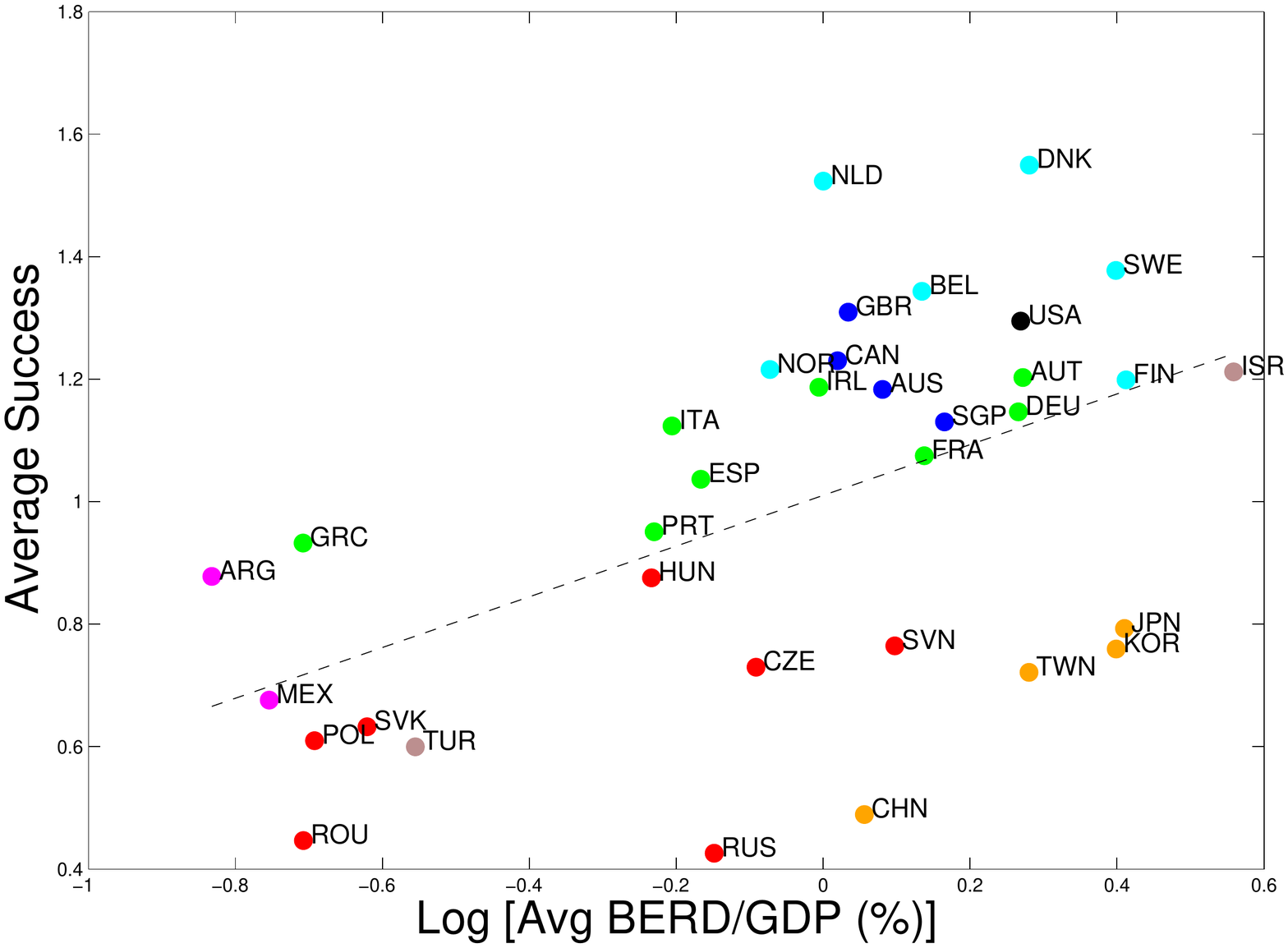}
\end{center}
\vspace{-0.5cm}
\caption{Relation between average BERD/GDP $\bar{\mathcal{B}}$ and average scientific success $\bar{\mathcal{S}}$ of nations.
Linear regression of data with slope $0.41\pm0.12$, intercept $1.01\pm0.05$ and $R^2=0.27$ is shown as a dashed segment.}
\label{fig:SuccBERD}
\end{figure}

\begin{figure}[p]
\begin{center}
\includegraphics[width=0.8\textwidth]{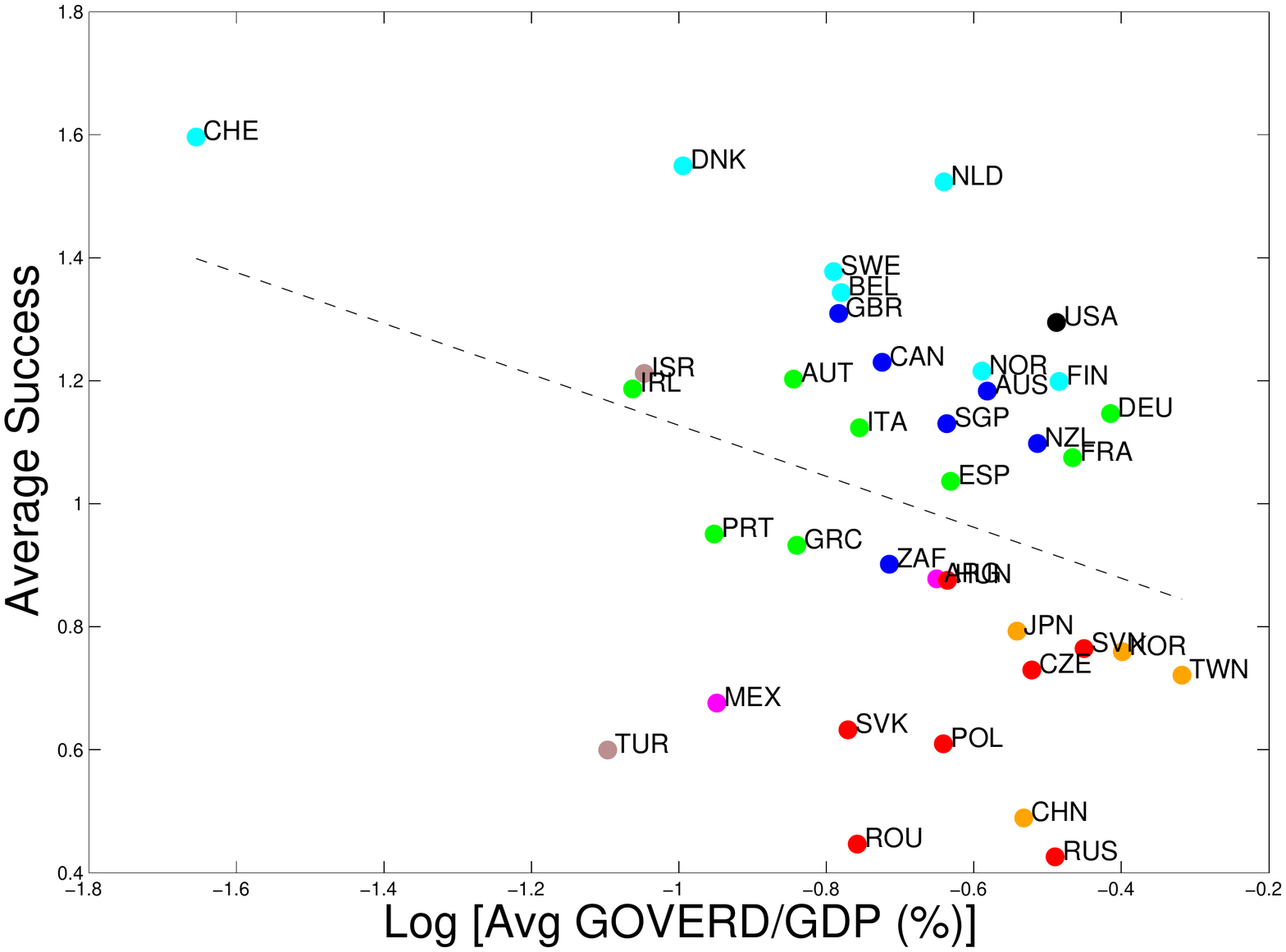}
\end{center}
\vspace{-0.5cm}
\caption{Relation between average GOVERD/GDP $\bar{\mathcal{G}}$ and average scientific success $\bar{\mathcal{S}}$ of nations.
Linear regression of data with slope $-0.41\pm0.20$, intercept $0.71\pm0.15$ and $R^2=0.11$ is shown as a dashed segment.}
\label{fig:SuccGOVerd}
\end{figure}

\begin{figure}[p]
\begin{center}
\includegraphics[width=0.8\textwidth]{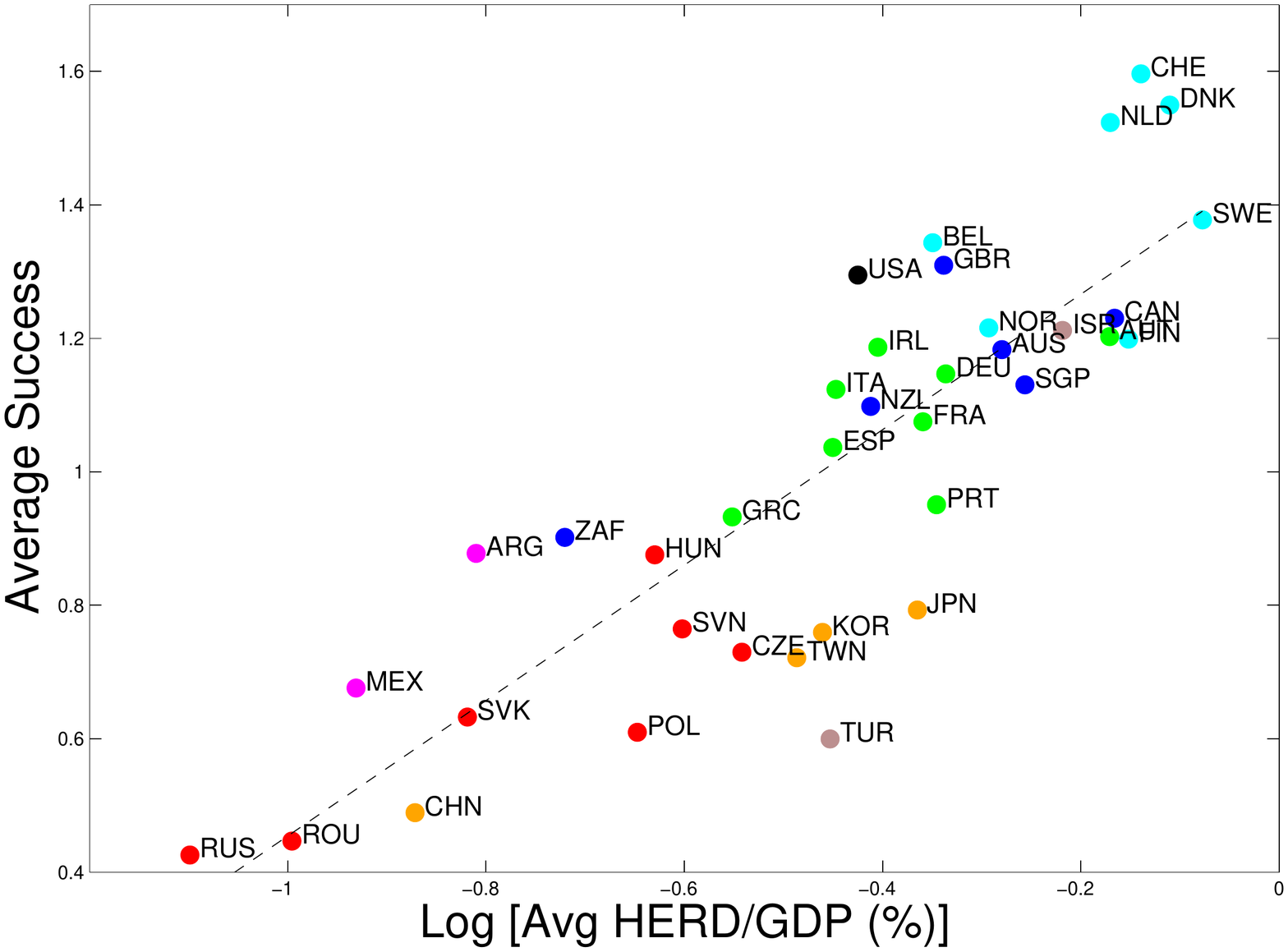}
\end{center}
\vspace{-0.5cm}
\caption{Relation between average HERD/GDP $\bar{\mathcal{H}}$ and average scientific success $\bar{\mathcal{S}}$ of nations.
Linear regression of data with slope $1.01\pm0.11$, intercept $1.47\pm0.06$ and $R^2=0.72$ is shown as a dashed segment.}
\label{fig:SuccHERD}
\end{figure}

\begin{figure}[p]
\begin{center}
\includegraphics[width=0.8\textwidth]{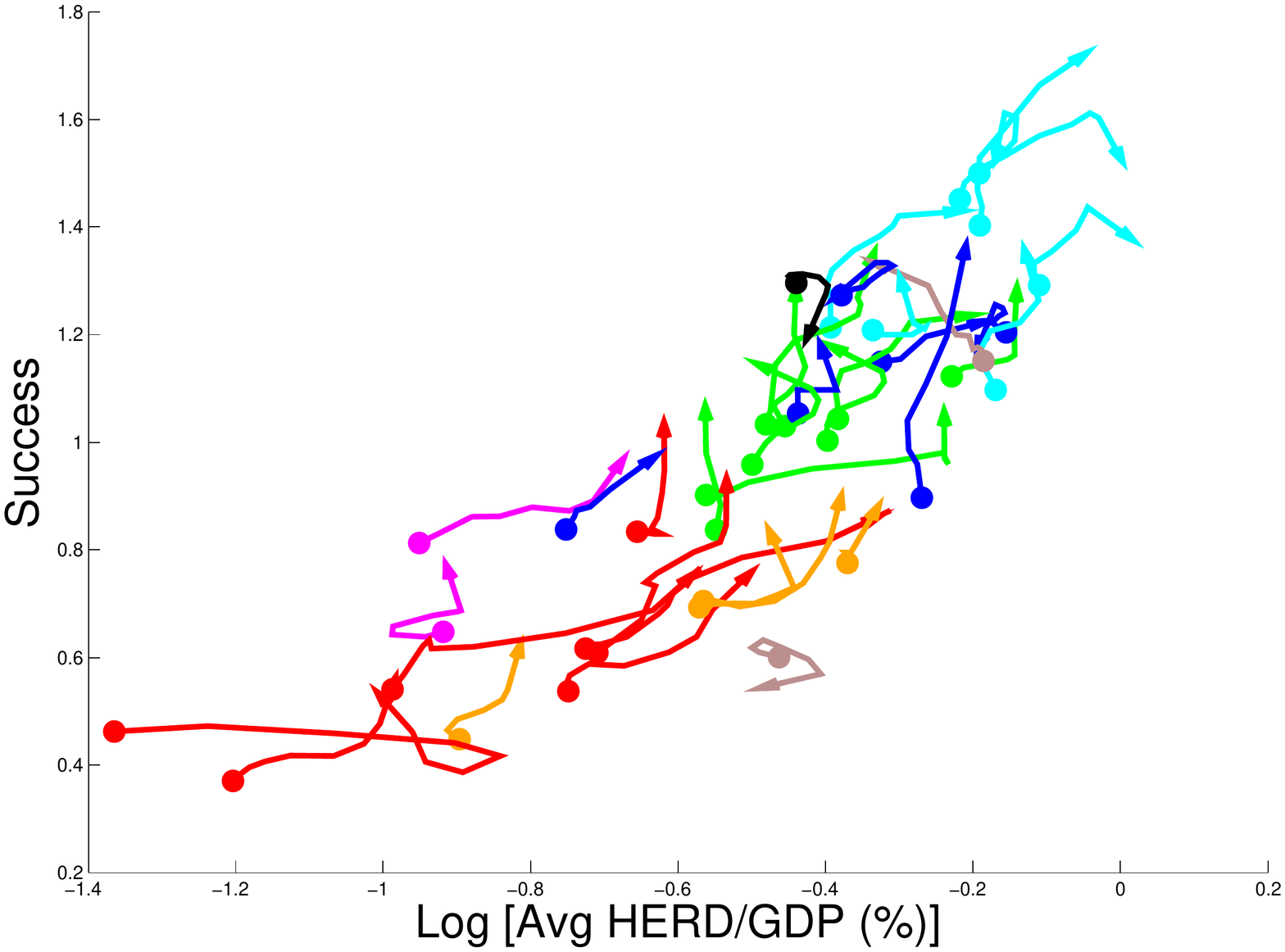}
\end{center}
\vspace{-0.5cm}
\caption{Temporal evolution of nations in the plane of HERD/GDP $\mathcal{H}(t)$ and scientific success $\mathcal{S}(t)$.}
\label{fig:SuccHERD_evol}
\end{figure}

We now turn our attention to the relation between ``input'' (represented by R\&D funding) and ``output'' (bibliometric-based success) of national research systems. 
For the reasons explained in section Methods, we consider all three indicators (BERD, HERD and GOVERD, normalized by GDP) as measures for national research expenditures. 
Among them, BERD is arguably more related to an output in patents than in publications, yet the correlation between these kinds of output is likely to be high: 
the most competitive nations in science are also the most competitive in technology, and several causal and feedback relations exist between the creation of knowledge and the development of complex products 
\citep{Cimini2014}. Indeed, figure \ref{fig:SuccBERD} shows that the scientific performance of nations is moderately correlated with BERD/GDP, and several patterns emerge. 
The main group, consisting of (most) Western and Northern Europe, Commonwealth members, United States and Israel, is characterized by high values of both scientific impact and R\&D business expenditure, 
with Italy, Spain and Portugal slightly falling behind in terms of $\bar{\mathcal{B}}$. Asian countries have similar BERD/GDP values to those of the western countries, but much lower scientific success. 
Eastern European countries are instead split into two groups, depending on their $\bar{\mathcal{B}}$ values. Finally, nations with the lowest BERD/GDP values are Argentina, Mexico and Greece, 
countries that underwent (or are currently undergoing) a sovereign debt crisis---that supposedly affected investors trust and thus private fundings.

On the other hand, as figure \ref{fig:SuccGOVerd} clearly shows, GOVERD/GDP is not related at all to the scientific performance of nations. This happens for two main reasons: 
i) research institutions that are internal to the government are generally less compelled to publish papers than research centers related to higher education, 
and therefore their output is aimed more at practical applications than at knowledge dissemination; as a consequence, bibliometric indicators cannot be suitable measures of their success; 
ii) GOVERD/GDP is a highly varying small percentage of GERD/GDP for most nations and thus more prone to noisy fluctuations than HERD/GDP: 
excluding peculiar countries like Russia and China, GOVERD amounts on average to 67\% of HERD---a percentage that decreases to 43\% when only western countries are considered. 
%Indeed, nations with the highest GOVERD over HERD ratio are the less successful ones according to our measures. 
Overall, this results in a correlation between impact and GOVERD/GDP that, excluding the significant outlier represented by Switzerland, 
amounts to $-0.2$: GOVERD is thus not informative to national scientific success.

Finally, and not surprisingly, HERD/GDP has the highest correlation with success, as shown by figures \ref{fig:SuccHERD} and \ref{fig:SuccHERD_evol}. 
In particular, the average values reported in figure \ref{fig:SuccHERD} follow a definite trend that starts from Eastern European nations (lowest $\bar{\mathcal{H}}$ and $\bar{\mathcal{S}}$), 
continues with Asiatic and Latin American countries and then with Western Europe and Commonwealth members, and ends with Northern Europe and Switzerland (highest $\bar{\mathcal{H}}$ and $\bar{\mathcal{S}}$). 
Note that while a country placed above/below the mean trend features a more/less efficient research system, there are no significant outliers. 
The dynamical analysis of the time evolution of nations in the $\mathcal{H}-\mathcal{S}$ plane shown in figure \ref{fig:SuccHERD_evol} reveals that countries trajectories are generally smooth. 
Overall, most of the developed countries are increasing their R\&D investments in time 
(even in periods of financial instability represented by the 2007-2009 financial crisis), a fact that brings to the increase of scientific performance---with some exceptions. 
The steepest growth of $\mathcal{S}(t)$ with respect to $\mathcal{H}(t)$ is observed for nations like Singapore, Italy, Greece and Hungary. For the latter European countries, 
the increasing success at constant investment can be at least partially explained by the drive of EU funding instruments. Instead, significant outliers are represented by United States and Turkey, 
for which both investments and success are decreasing (at least during the latest time window), and Israel, whose scientific performance is increasing in spite of a decrease in funding. 
We recall again that decreasing of success for some countries is due to the separate conservation of shares in the measure of success.

Because of the highest correlation between ``inputs'' and ``outputs'' of research observed in this latter case, in the following we focus solely on HERD as a metric for R\&D funding. 
Differently form other works \citep{Zhou2006}, we prefer HERD/GDP to (HERD+GOVERD)/GDP as the highest correlation with success is observed for the former case 
(because of the different focus and noisy features of GOVERD we discussed above). 

\subsection{Emergence of geo-cultural clusters}

\begin{figure}[h!]
\begin{center}
\includegraphics[width=0.6\textwidth,angle=-90]{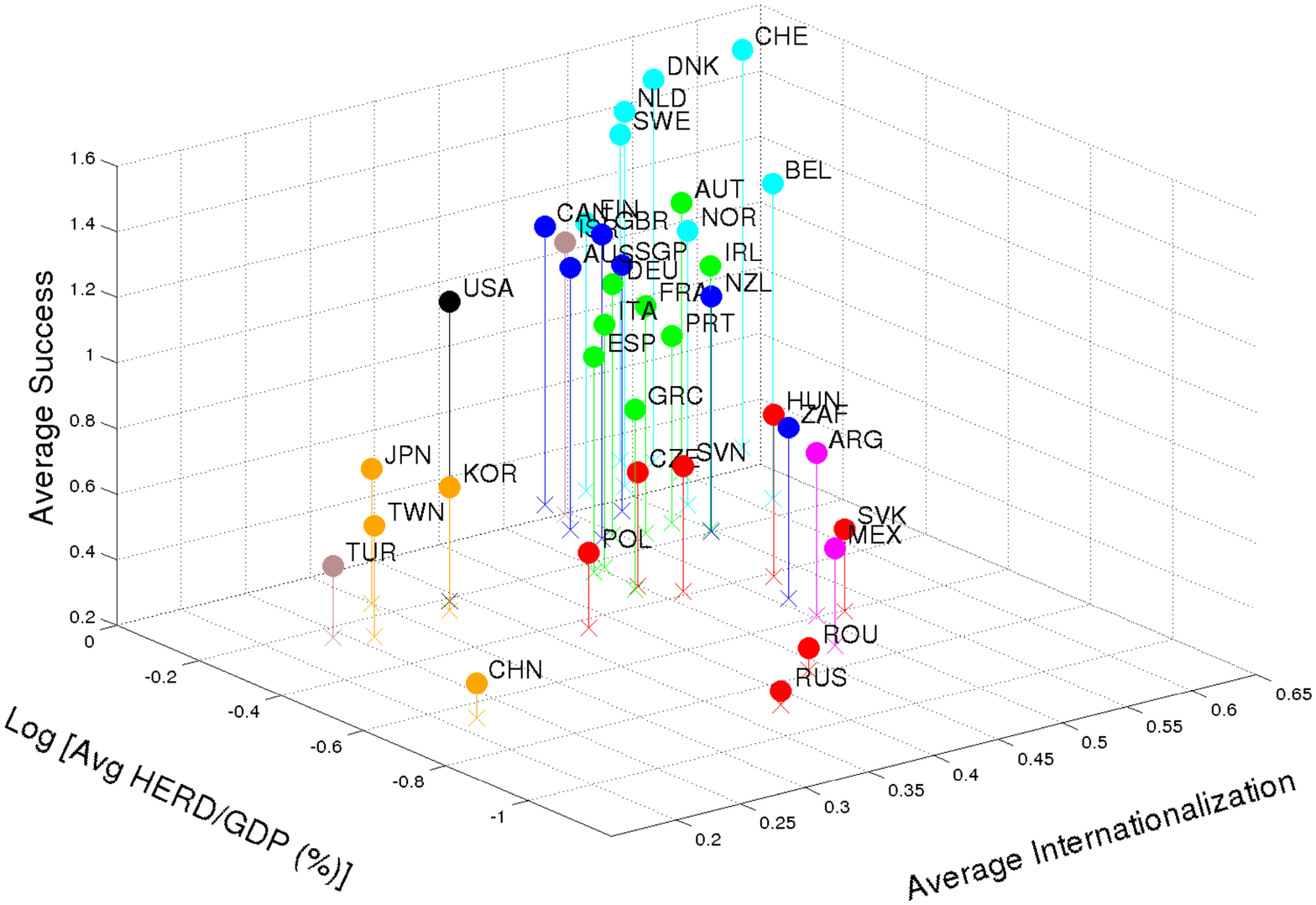}
\end{center}
\vspace{-0.5cm}
\caption{Relation between average HERD/GDP $\bar{\mathcal{H}}$, average internationalization $\bar{\mathcal{I}}$ and average scientific success $\bar{\mathcal{S}}$ of nations.}
\label{fig:3D}
\end{figure}

We now put together the three fundamental features of national research system, meaning that we analyse the position of individual nations in the three dimensional space 
defined by $\mathcal{I}$, $\mathcal{H}$ and $\mathcal{S}$. Figure \ref{fig:3D} shows that an even more clear geo-cultural structure naturally emerges in this space. 
The group of Asian nations is clustered together around medium-low values of HERD/GDP and very low internationalization, with China still behind the other countries in terms of scientific success. 
United States have similar values of $\mathcal{I}$ and $\mathcal{H}$, but they are disconnected from the group of Asian countries because of their much higher performance. 
East European nations are instead located (together with Latin America and South Africa) in the region of very low HERD/GDP, low internationalization and---as expected from the previous analysis---low success. 
Concerning in particular the East European countries, a gradual detaching pattern from Russia emerges: while Russia seems not to be recovering from the radical drop of investments 
since the break-up of the Soviet Union, the other nations are gaining scientific success by increasing R\&D investments. 
However, while in some cases (Poland, Czech Republic, Slovenia) HERD/GDP values are now comparable to those of western countries, the scientific performance of ex-soviet influenced countries is still low---a fact 
that questions the effectiveness of EU fundings in the East Europe area up to now. Moving further, the central dense cluster is composed of Western European countries and Commonwealth members 
(the latter having slightly higher HERD/GDP and success). Finally, Northern Europe and Switzerland are located in the region where all the fundamentals assume their highest values.
Notably, these clusters are coherent with those found by \citet{Frame1979} in the late `70 , and by \citet{Luukkonen1992} in the early `90: 
they are inherited from the past, and seem indeed to endure in spite of the increasing internationalization of science and the integration policy of the EU in particular. 
Yet, these past works focused mainly on international co-authorships. By adding research investments and success as fundamental characterizing features, 
our analysis strengthen the importance of extra-scientific factors (history, geography, politics and language) in shaping the structure of national research systems.

\section{Conclusions}\label{sec:conclusions}

The ability to assess the impact of the scientific system of a nation is crucial for both public and private stakeholders to determine scientific priorities and investments \citep{May1997,King2004}. 
In this work, we have characterized national research systems through three fundamental features: R\&D investments, internationalization and bibliometric performance. 
We have systematically studied the evolution of nations in the space of these variables, and discussed the emerging patterns of geo-cultural affinities between nations. 

In the context of assessing the national scientific impact, we have employed recently proposed approaches based on ratios of citations to publications, 
and discussed some variants of these metrics that, remarkably, are independent on the specific classification used for scientific sectors. 
Note that whatever the precise matematical definition, all these indicators reward nations with high number of citations per paper, and are therefore intensive. 
This is the reason why countries (like China) having a rapid growth in publication output, but only a gradual improvement of citation impact, do not perform well according to our study. 

Concerning research investments, we have focused on HERD (basically, the expenditure to form and fund highly qualified research personnel at universities), 
showing that government intramural fundings (GOVERD) provide basically no information on the quality of scientific output, whereas, 
R\&D expenses in the business sector (BERD) do provide some information but because of the correlation between scientific and technological performance in developed countries. 
Thus, overcoming the ambiguity of R\&D funding schemes by choosing the appropriate variable, we have confirmed that the scientific impact of a nation grows with the amount of funding. 
Yet, it is important to remark that there are lags between changes in research funding and publication outputs (as for outputs and their impact) of the order of two-three years, 
so that multivariate models may be necessary to deeply understand the productive dynamics of science and innovation \citep{King2004}. 

We have also confirmed the findings of previous studies that the amount of international collaboration in science is steadily growing in time, 
and that almost all nations are nowadays involved in international collaborations---which generally lead to research of higher impact. 
Internationalization is increasing especially in Western and Northern Europe, 
supposedly as an effect of EU Commission's efforts to stimulate collaboration within European countries. 
The fact that EU countries are also increasing scientific success to top levels and at fast peace testifies the rate of progress towards the European Research Area (ERA) vision. 
Note that international collaborations are especially important for fostering frontier research, needed to address global challenges that requires input from a wide range of expertises. 
However, according to {\em Science Europe} \citep{SciEU}, with the European Research Council (ERC) grant scheme being dedicated to investigator-driven research, the current absence of ERC Synergy Grants 
from the funding schemes, and the Horizon 2020 Societal Challenges primarily focused on near-market applied research, funding opportunities for collaborative basic and frontier research at the EU level 
are at the moment quite limited. This may put future scientific progress at a stake. Nevertheless, internationalization emerges from our analysis as an additional fundamental parameter 
for the scientific development of nations, and in this sense the EU has been rather successful (up to now).

Putting together all three fundamentals, clearly discernible patterns do emerge, caused by various factors as different as history, geography, politics, language. 
For instance, Eastern Europe is slowly detaching from the Russian attractor, trying to catch up with the rest of EU nations that have a remarkable performance. 
In particular, Northern European countries and Switzerland are the top players according to the intensive metrics we use---that defines 
the mean quality of single papers rather than the overall national scientific impact.
On the other hand, the Asian nations are not increasing the level of internationalization of their research systems, possibly because countries like Japan and China are more scientifically isolated 
than other developed countries. Additionally, large research systems rely less on international co-authorship, as also pointed out by \citet{Frame1979}. 
This is testified by the case of United States, that evolve similarly to western counties in internationalization but with a large negative offset due to their larger self-consistency. 
Indeed, relatively small-sized and geographically close countries like European nations are facilitated in the establishment of cross-border international collaborations with respect to, e.g., 
USA and China. Note however that while the absolute values of internationalization and impact we obtain are affected by this phenomenon, and are also due to the counting method used to assign publications 
to countries (see section Methods above), the relative changes emerging from the temporal dynamics of these variables are not. This clearly testifies the effects of EU integration policies 
pushing for the establishment and reinforcement of a stable European collaboration network. 
Yet, in order to properly quantify these effects, more detailed analyses are needed. For instance, the USA and China could be compared with the EU as a whole, or, alternatively, 
each state of the USA and each province of China could be considered as an independent nation. Note however that achieving a fair comparison at this level is difficult, 
because important factors like languages and internal regulations are quite different between the various European countries, whereas, for USA and China they are the same. 
Nevertheless, these kind of studies can be of significant interest from a policy perspective, and will be the subjects of future research.

Finally, note that our assessment of the scientific success of nations based on publication impact is in line with the quantification of the level of scientific diversification and competitiveness, 
pursued through appropriate algorithm that leverage on the detailed structure of national research systems \citep{Cimini2014}. 
Remarkably, this latter approach was originally developed for economics, and used to successfully measure the economic potential and competitiveness of nations, together with the complexity of produced goods 
\citep{Caldarelli2012,Tacchella2013}. Scientific and economic production of nations thus seems to follow similar structural patterns. 
Moreover, the heterogeneous dynamics of nations we find here reflect those found for economic development (laminar for developed countries, chaotic for underdeveloped countries) \citep{Cristelli2015}. 
An interesting perspective thus could be the study of scientific development in the context of diffusive dynamics of nations in the space of scientific domains, 
as it was done for economic development in the space of products \citep{Zaccaria2014}. Overall, the parallelism found between scientific and economic production 
can be seen as a natural consequence of the coupling and co-evolution of the different compartments of the innovation ecosystem. 
We believe future research will be bound to face the challenge of identifying the micro-determinants and their complex interactions 
that are responsible for the observed emergent macro-properties of the innovation ecosystem, 
allowing to unfold the complex interplay between scientific advancement, technological progress, economic development and societal changes.

\section*{Acknowledgments}
This work was supported by the EU projects GROWTHCOM (FP7-ICT, grant n. 611272), MULTIPLEX (FET, grant n. 317532), SIMPOL (FP7-ICT, grant n. 610704), CoeGSS (EINFRA, n. 676547), and the Italian PNR project CRISIS-Lab. 
The funders had no role in study design, data collection and analysis, decision to publish, or preparation of the manuscript.

\end{document}